\documentclass[manuscript=article,nalefd]{achemso}

\usepackage{amssymb} 
\usepackage{graphicx}
\usepackage{dcolumn}
\usepackage{bm}
\usepackage{siunitx}
\usepackage[percent]{overpic}
 \usepackage{tikz}
\newcommand{\virgolette}[1]{``#1''}

\usepackage{comment}
\usepackage{color}
\usepackage{xcolor}
\newcommand{\corr}{\color{black} }

\usepackage{pict2e} 
\usepackage{float}
\usepackage{braket}

\usepackage[export]{adjustbox}

\usepackage[normalem]{ulem} 

\usepackage[T1]{fontenc} 
\usepackage{amsmath}
 \usepackage{hyperref}
\hypersetup{
    colorlinks,
    linkcolor={blue!80!black},
    citecolor={blue!50!black},
    urlcolor={blue!50!black}
}

 \newcommand{\MD}{\textcolor{black}} 
  
 \newcommand{\AB}{\textcolor{black}}

\makeatletter
\def\@email#1#2{%
 \endgroup
 \patchcmd{\titleblock@produce}
  {\frontmatter@RRAPformat}
  {\frontmatter@RRAPformat{\produce@RRAP{*#1\href{mailto:#2}{#2}}}\frontmatter@RRAPformat}
  {}{}
}%
\makeatother

\title{
Hidden vibrational bistability revealed by intrinsic fluctuations of a carbon nanotube
} 

\author{P. Belardinelli}
 \affiliation{Department of Construction, Civil Engineering and Architecture, Polytechnic University of Marche, 60131 Ancona, Italy}%
 \author{W. Yang}\affiliation{ICFO - Institut de Ciencies Fotoniques, The Barcelona Institute of Science and Technology, 08860 Castelldefels, Barcelona, Spain}%
\author{A. Bachtold}\email{Adrian.Bachtold@icfo.eu}%
\affiliation{ICFO - Institut de Ciencies Fotoniques, The Barcelona Institute of Science and Technology, 08860 Castelldefels, Barcelona, Spain}%
\author{M.I. Dykman}\email{dykmanm@msu.edu}%
\affiliation{Department of Physics and Astronomy, Michigan State University, East Lansing, MI 48824, USA}%
\author{F. Alijani}\email{f.alijani@tudelft.nl}%
\affiliation{Department of Precision and Microsystems Engineering, Delft University of Technology, Mekelweg 2, 2628CD, Delft}%

\begin{document}
\begin{abstract} 
We demonstrate that a quiet state and large-amplitude self-sustained oscillations can co-exist in a carbon nanotube subject to time-independent drive. \MD{A feature of the bistability is that it} would be hysteresis-free in the absence of noise \MD{and the oscillatory state would not be seen.} It is revealed by \MD{ random switching between the stable states, which we observe in the time domain.}   We attribute the switching to   fluctuations \MD{in the system} and show that  it displays Poisson statistics. We propose a \MD{minimalistic} model   that \MD{relates}  the emergence of \MD{the} bistability to a {\corr non-monotonic} variation of nonlinear friction with the vibration amplitude. This new type of dynamical regime and the means to reveal it are generic and are of interest for various mesoscopic vibrational systems.
\end{abstract} 

\maketitle 

{\corr
}\\
Nano-electro-mechanical systems (NEMS) provide a means for studying physics away from thermal equilibrium in a well-characterized setting \cite{Bachtold2022a}. 
 An important group of non-equilibrium phenomena originates from the 
 interplay between nonlinearity and fluctuations in driven systems, 
 which can modify the {\corr 
 frequency stability \cite{Greywall1994,Villanueva2013,Kenig2013,ZhangYan2024,Sadeghi2020},} 
 the  power spectrum \cite{Stambaugh2006a,Huber2020},  lead to {\corr spectral broadening \cite{Rechnitz2022}}, and thermal noise squeezing \cite{Buks2006,Huber2020,Yang2021b}. The interplay  is most nontrivial when a non-equilibrium system is brought into a regime where it exhibits bistability. 
 Here, fluctuations, even if weak on average, can cause interstate transitions and are ultimately responsible for the distribution of a system over the stable states\cite{Bachtold2022a}. Much work on studying these effects and the emerging scaling \cite{Dykman1980,Dykman2007} has been carried out on nano- and micromechanical resonators driven by an external  resonant   force or modulated parametrically \cite{Aldridge2005,Stambaugh2006a,Chan2007,Karabalin2011,Venstra2013,Defoort2015,Chowdhury2017,Dolleman2019,Steele2009,Lassagne2009,Tabanerabravo2022}. 
 A mechanism that leads to the onset of bistability in NEMS without periodic driving was suggested in Ref.~\cite{Usmani2007} and such bistability was observed in a carbon nanotube (CNT) \cite{Wen2020}.

In almost all bistable mesoscopic vibrational systems,  the vibrations could be brought to one of the stable states by smoothly changing a control parameter, for example, the driving force. As a result of the change, at some critical parameter value, the bifurcation point, one of the stable states would lose stability, and the system would switch to another stable state.  Such behavior  is usually accompanied by hysteresis: in a parameter range between bifurcation points the state of the system depends on the history of the parameter change.  In all  works on NEMS thus far \MD{hysteresis was used to reveal the bistability.}  

\begin{figure}[htb!]
\centering \includegraphics[width=1\textwidth]{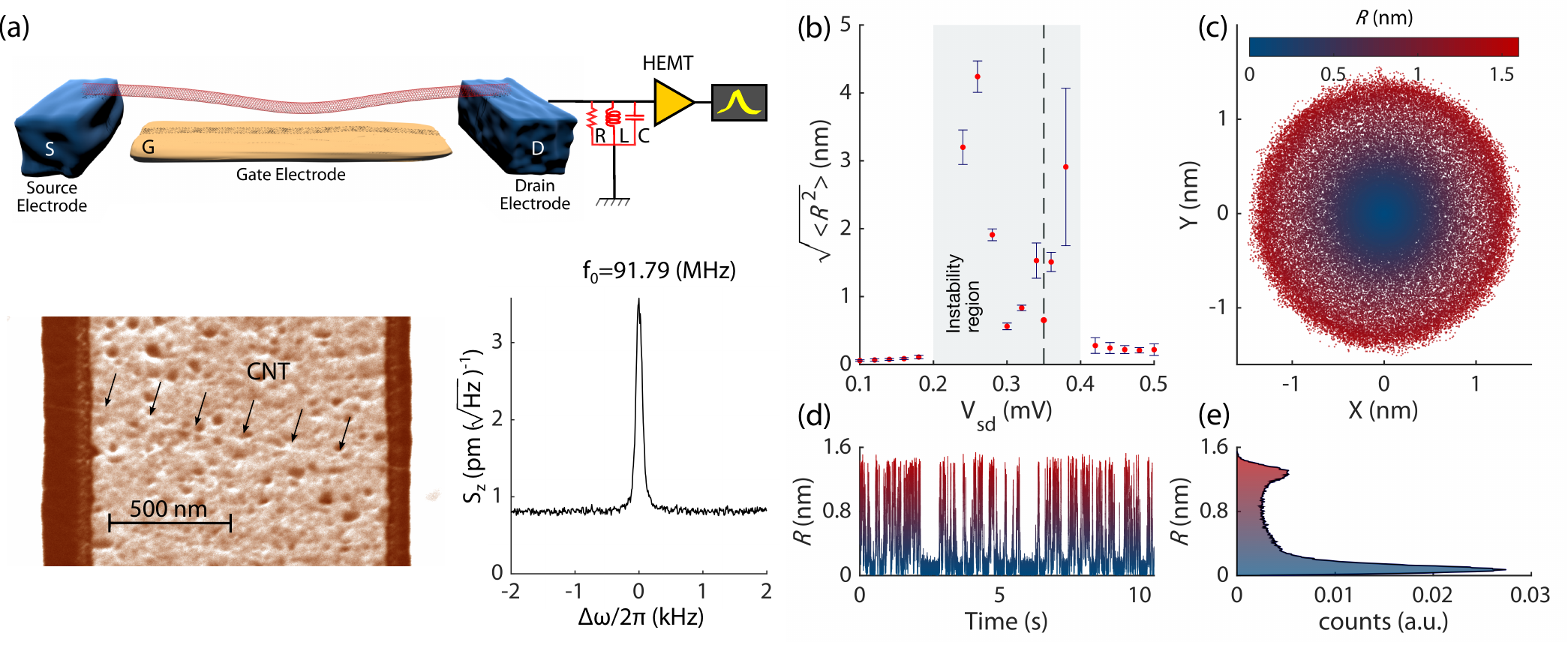}
\caption{ \corr (a) CNT-based electromechanical oscillator and measurement schematic.   The CNT has the length of $\approx 1.5~\mu$m and the radius of $\approx 1$~nm (scanning electron microscope image in the bottom left panel). Voltages $V_\text{sd}$ and $V_\text{g}$ are applied to electrodes S and Gate, respectively. The drain electrode D is connected to an RLC resonator ($f_\text{RLC}$=1.27 MHz). The displacement spectral density is shown in the bottom right panel.  
 (b) The   root mean square nanotube displacement from the origin   $\sqrt{<R^2>}$ as a function of the source-drain voltage $V_\text{sd}$ for  $V_\text{g}=$\SI{-616}{\milli\volt};  $\sqrt{<R^2>}$ is obtained from spectral noise measurements for all the data points except for $V_\text{sd}= $\SI{0.35}{\milli\volt} (dashed line), the latter is obtained by recording the time evolution of the displacement amplitude at the mode antinode. (c)   The measured two quadratures $\text{X}$ and $\text{Y}$ of the motion on the $(\text{X},\text{Y})$-plane at this $V_\mathrm{sd}$, (d) fluctuations of the amplitude $R{}=\sqrt{\text{X}^2+\text{Y}^2}$ in time for the same $V_\mathrm{sd}$,  
 (e) the amplitude  histogram, normalized with respect to the total number of observations.
 }
\label{fig:observation} 
\end{figure}

In this paper, we \MD{report the observation of  a  hysteresis-free bistability in  a nanomechanical system. Such bistability means that, as the the control parameter is changed back and forth, the system remains in the quiet state. The very presence of another stable state is revealed by fluctuations that cause interstate transitions. In our system,}  
coexisting are   the quiet state and \MD{the state of} large-amplitude self-sustained oscillations of the lowest mode of a CNT driven by a time-independent source-drain voltage, see Fig.~\ref{fig:observation}.  The large-amplitude oscillatory state was already identified in Ref.~\cite{Urgell2020}. \MD{It is important that the mode experiences fluctuations.
We reveal that the system is actually bistable by examining the fluctuation statistics. In distinction from the more conventional scenario, } the large-amplitude oscillatory state emerges as the source-drain voltage is increased, and when it emerges, it is already well separated from  the quiet state in phase space.  \MD{Then,} as the source-drain voltage is further changed, the  vibrational  state  is observed to  lose stability. 
There is no hysteresis when the bifurcation parameter is moved back and forth. In terms of the bifurcation theory \cite{Dellwo1982},
the emerging and disappearing stable states are associated with an \virgolette{isola}: {\corr an isolated branch of an equilibrium state of a noise-free system.} 

 The experiment is done using clamped-clamped CNTs grown by chemical vapour deposition across two metallic contact electrodes \cite{Moser2014}. The measurements are  performed at cryogenic temperature (70mK) by applying a voltage bias to the source  ($V_\text{sd}$) and to the gate electrode ($V_g$) which is placed beneath  the CNT. The current from the drain electrode is measured by using a RLC resonant circuit  
 and a low-temperature HEMT amplifier, see Fig.~\ref{fig:observation}(a). 
 The read-out signal is obtained by measuring the 
 current noise spectrum, which is converted to the nanotube displacement  \cite{Urgell2020}. The same calibration is used to obtain   the quadratures of the motion (X,Y) from the lock-in measurements.  
 The scaled vibration amplitude is $R{}=\sqrt{\text{X}^2+\text{Y}^2}$.

When sweeping up the source-drain voltage $V_\text{sd}$ at $V_\text{g}=$\SI{-616}{\milli\volt}, there occurs a sudden jump up in the displacement $R$ for  $V_\text{sd}\approx $\SI{0.2}{\milli\volt}  followed by a jump down at $V_\text{sd}\approx $\SI{0.4}{\milli\volt}, see  highlighted region of Fig.~\ref{fig:observation}(b). The change of the nanotube motion with $V_\text{sd}$ in Fig.~\ref{fig:observation}(b) does not resemble that of a vibrational system undergoing \MD{a supercritical} Hopf bifurcation, whose signature is a smooth monotonic increment of the oscillation amplitude  and loss of stability of the quiet state \cite{Zakharova2010}. The instability is observed over a narrow range of gate voltage $V_\text{g}$ but reappears periodically in $V_\text{g}$ with a period corresponding to adding four electrons to the nanotube.  
{\corr The periodicity is consistent with the SU(4) symmetry of the CNT \cite{Urgell2020}.}

In the (X,Y) space  at $V_\text{sd}= $\SI{0.35}{\milli\volt} we recognize a \MD{highly populated} doughnut-like \MD{region centered at}  a nonzero mean amplitude that 
encircles the thermal motion about the origin of the quadrature space. \MD{This region suggests}   the presence of an oscillatory state, see  Fig.~\ref{fig:observation}(c). 
%
The time trace of the motion (Fig.~\ref{fig:observation}(d))  and the normalized histogram of the amplitude $R{}$ in Fig.~\ref{fig:observation}(e) \MD{further support the notion}   that the system has two different dynamical states. 
While distinct peaks in the amplitude distribution may serve as a signature of co-existing states, a double-peak pattern can also arise in various other dynamical phenomena such as intermittent chaos or bursting oscillations \cite{nayfeh2008applied}.  A careful  analysis is required   to differentiate bistability from aperiodic dynamical behaviors.

\begin{figure}[htb!]
\centering    
\includegraphics[width=0.65\textwidth]{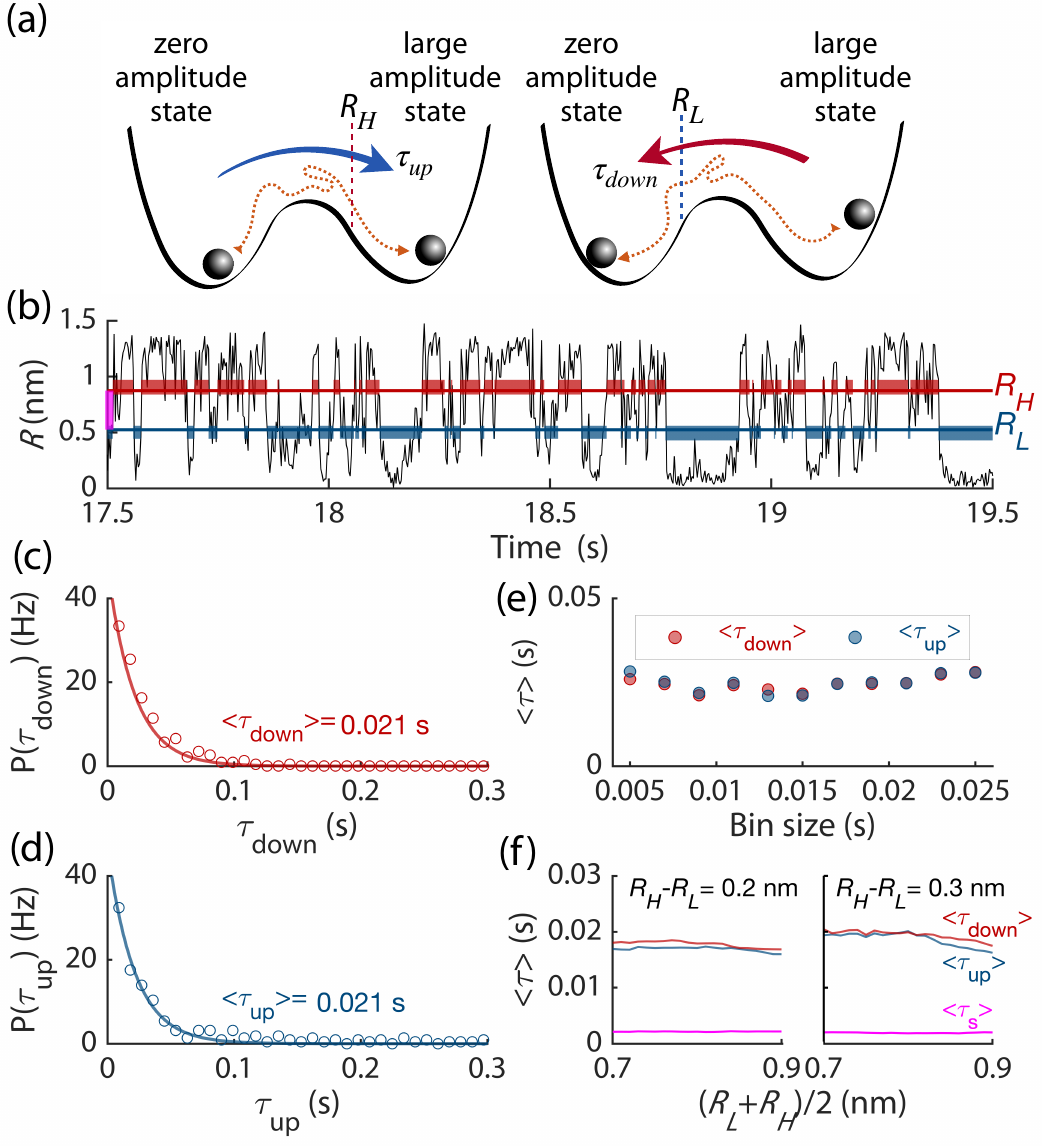}
\caption{
  {\corr (a) 
  A double-well potential, illustrating bistable dynamics using a ball-in-a-cup analogy.} 
  The minima are associated with the stable state of self-sustained vibrations and the zero-amplitude state of the CNT. Noise-induced transitions to the large-amplitude (zero-amplitude) states are considered to occur once the vibration amplitude crosses the threshold $R_H (R_L)$.  
  (b) A sample of the time evolution of the  vibration amplitude  ($V_\text{sd}= $\SI{0.35}{\milli\volt}). 
  Blue/red bars indicate the chosen switching thresholds  with R$_{H}$-R$_{L}$=0.35nm, 
 (R$_{H}$+R$_{L}$)/2=0.7nm. The magenta bar indicates the region between $R_H$ and $R_L$. 
  (c)-(d) Dwell (residence) time distributions (bin size 9 ms)  for the large-amplitude state [panel (c)] and   the zero-amplitude state [panel (d)].  A  Poisson distribution (Eq.~\eqref{eq:poisson})  is fitted to the data.  It   gives averaged dwell times of $<\tau_\text{down}>=0.021$ s and $<\tau_\text{up}>=0.021$ s. (e) Influence of the bin size on the average dwell times in panels c and d. (f) Average dwell times for varying thresholds $R_{L}$ and $R_{H}$. The average time $<\tau_{s}>$ is the time spent in between the two thresholds.
 }
\label{fig:switching} 
\end{figure}

To understand the nature of  the observed dynamics,  we perform statistical analysis on the time-domain data of Fig.~\ref{fig:observation}(d).    \MD{This dataset has not been presented  in the earlier work} \cite{Urgell2020}.
We assume  that our system has two stable  states, {\corr namely a zero-amplitude, i.e. a quite state, and  a self-sustained oscillatory state, which has a large  amplitude compared to the root-mean-square amplitude fluctuations \cite{Willick2020,Weldon2010,Feng2008,Sekaric2002}}. 
%
We   investigate whether noise   induces stochastic transitions \MD{between the stable states},   analogous to the noise-induced \MD{interwell} hopping of a damped particle  in a  double-well potential, see  Fig.~\ref{fig:switching}(a), \MD{cf. \cite{Zhang2024} ; such hopping underlies stochastic resonance \cite{Benzi1981,Nicolis1982}.} 
The difference \MD{in our case}  is that 
here one of the stable states is a static equilibrium point, whereas the other is a state of self-sustained vibrations.

Noise-induced switching is well-defined if the switching rate is much smaller than the relaxation rate. A noise-driven system then spends most of the time fluctuating about one of its stable states. The characteristic correlation time $t_r$ of these fluctuations is the dynamical relaxation time or the correlation time of the noise. Occasionally there occur large outbursts of noise that lead to switching between the states. The typical time between such outbursts is much larger than $t_r$, whereas the duration of the switching event itself is comparable to $t_r$. Therefore the switching events are expected to be uncorrelated and described by the Poissonian statistics.

In order to detect interstate switching events, one would need to set a threshold, which is related to but does not coincide with the basin boundary of the two states. 
For a particle in a double-well potential,  
reaching the barrier top does not necessarily lead to switching\MD{, as the system can go back to the initially occupied well}. Even in the simplest case of fluctuations induced by white noise, to find the switching rate one has to set up a threshold sufficiently far beyond the barrier top with respect to the initially occupied state, as has been known since the classical work of Kramers \cite{Kramers1940}. It is clear from the above arguments that the thresholds for transitions between different states should not coincide, cf. \cite{Dykman1991}.  
 Hence,  we introduce two amplitude thresholds, $R{}_{L}$ and $R_{H}$, see Fig.~\ref{fig:switching}(a). If the system was fluctuating about $R{}=0$   and its trajectory   crossed $R{}_{H}$, we assume that it has switched to the large-amplitude state. On the other hand, if the system was fluctuating about the large-amplitude state and its trajectory $R{}(t)$ crossed $R{}_{L}$, it has switched to the $R{}=0$-state. 
With the above definition,  the dwell (residence) times $\tau_\text{up}$ and  $\tau_\text{down}$   are the times spent in the zero- and large-amplitude states, respectively, before the system switches.  
The time intervals $\tau_\mathrm{up}$ and $\tau_\mathrm{down}$ 
are  shown in Fig~\ref{fig:switching}(b) by the red and blue thick bars, respectively. 
The region between $R{}_{H}$ and $R{}_{L}$ 
contains the separatrix. The experimental data do not allow us to find it.

Figures~\ref{fig:switching}(c) and \ref{fig:switching}(d) show the distribution of the dwell times $\tau_\mathrm{up}$ and $\tau_\mathrm{down}$. 
The transitions between the states are well described by a Poisson process, and the distribution of the dwell times is close to exponential,  
 \begin{equation}
     P(\tau)=\frac{1}{<\tau>}e^{-\tau/<\tau>}.
     \label{eq:poisson}
 \end{equation}  
\MD{This is the central argument in support of the coexistence of two stable states in our system.} 
We use Eq.~(\ref{eq:poisson}) to fit the experimental data and find that the dwell times 
are approximately the same for the chosen parameters, with  
 $\tau_\text{up}\approx \tau_\mathrm{down} \approx$~21 ms. The fit is only mildly influenced by the bin size, see Fig.~\ref{fig:switching}(e).
These dwell times  are  much larger than the relaxation time $t_r$ 
of the nanotube, which can be inferred from duration of the switching events themselves:  in Fig.~\ref{fig:switching}(b) the trajectories leading to transitions are essentially vertical. 
The detailed data indicates that  $t_r \sim$ 
1-3 ms, as illustrated in    Fig.~\ref{fig:switching}(f) by the
average time $<\tau_{s}>$  spent in between the two
thresholds.  

\MD{Another important argument in support of the bistability is seen from}   Fig.~\ref{fig:switching}(f). \MD{In this figure} we  plot the   dwell times over  a broad range of mean threshold values and separations. 
The results do not change.  This demonstrates the reliability and stability of the two-threshold approach and confirms  the presence of  noise-induced hopping between  two metastable states.  
The stochastic analysis conducted on an additional temporal dataset, corresponding to $V_\text{sd}= $\SI{0.25}{\milli\volt}, 
 also aligns with these findings  (see   Supporting Information ~S1 \cite{SI}). 

 The bistable dynamics observed in Fig.~\ref{fig:switching}  ultimately comes from the  source-drain  voltage  $V_\mathrm{sd}$\MD{, which pumps energy into the system. The onset of self-sustained vibrations due to energy pumping is often associated with the friction coefficient becoming negative, which makes the quiet state unstable. In contrast, in our system the quiet state remains stable. This can be understood if the friction coefficient becomes negative in a certain range of sufficiently large vibration amplitude. } 
\MD{The dependence of the friction coefficient on amplitude is called nonlinear friction.} Such  friction is well-known in nanomechanics 
\cite{Keskekler2021,Bachtold2022a}. Usually it leads to a faster decay of vibrations with the increasing amplitude, that is, the coefficient of nonlinear friction is positive,  although there has been also observed slowing down of the decay with the increasing amplitude \cite{Singh2016}. 

\MD{There are several possible causes of nonlinear friction in our system.   One of them is the electron-vibrational coupling. As electrons hop between the leads and the nanoresonator, they exchange energy with the mode. This leads to decay or excitation of the vibrations, i.e., to positive or negative friction, see \cite{Armour2004,Novotny2003,Fedorets2004,Mozyrsky2006,Usmani2007,Samanta2023}  
and references therein. 
The analyses in these papers refer to the limit of strong Coulomb blockade. Negative nonlinear friction 
resulted from the dependence of the tunneling on the vibration amplitude \cite{Fedorets2004,Usmani2007}. 
This dependence should occur in our system, too, even though the Coulomb gap is moderately hard. One can picture this dependence as coming from the change of the transmission of the tunneling barrier due to the strain induced by the CNT displacement \cite{Minot2003}.
}

\MD{
In the basic model of the effect of vibrations on tunneling \cite{Kagan1992}
the vibration-induced change of the tunneling exponent is $C_\mathrm{tun} q/\lambda_\mathrm{tun}$, where $q$ is the mode coordinate and $\lambda_\mathrm{tun}$ is the electron tunneling length. \AB{In the measurements presented in this work, the} tunnel barriers are  defined in the clamping areas at the interface between the nanotube and the metal electrodes. For tunneling onto/from a CNT, the coefficient $C_\mathrm{tun}$ depends on the structure of this interface, which is not well characterized, and therefore it  cannot be found quantitatively.  However, the ratio $q/\lambda_\mathrm{tun}$ itself is $\gtrsim 10$ for the observed limit cycle radius and $\lambda_\mathrm{tun} \sim 3-4$~\AA. This suggests that the friction that results from the modulation of the tunneling barrier  can be significantly nonlinear. It can be negative in the range of $V_\mathrm{sd}$ where the energy transfer to the mode exceeds the energy drain from the mode. The amplitude dependence of the friction is affected also by the polaronic effect: because of the gate voltage, electrons exert force on the mode that depends on the number of electrons on the CNT, which itself depends on the CNT displacement. This force leads to vibration decay, for strong Coulomb blockade \cite{Mozyrsky2006}.
}

\MD{Another source of negative nonlinear friction, a  retarded backaction from the circuit, is discussed in the SI Sec.~S2\cite{SI}. Here, a quantitative theory requires the knowledge of the nonlinear dependence of the conductance on the vibration amplitude, which itself requires full characterization of the clamping area. However,  the magnitude of this nonlinear friction force compared to the linear one is proportional to the square of the ratio of the displacement amplitude to the distance to the gate electrode and thus relatively small.}

\MD{The above arguments show that the situation with nonlinear friction is no different from  linear friction, which comes from several possible mechanisms. Therefore, to describe the central experimental observation,  the onset and collapse of self-sustained vibrations, we use a minimalistic model. A major feature of this model is the absence of delay in the friction force, in the rotating frame. This is a consequence of the smoothness of the density of states of the electrons and thermal acoustic phonons involved in the mode decay processes, cf. \cite{Bachtold2022a} 
}

\begin{figure}[htb!]
\centering
\includegraphics[width=0.65\textwidth]{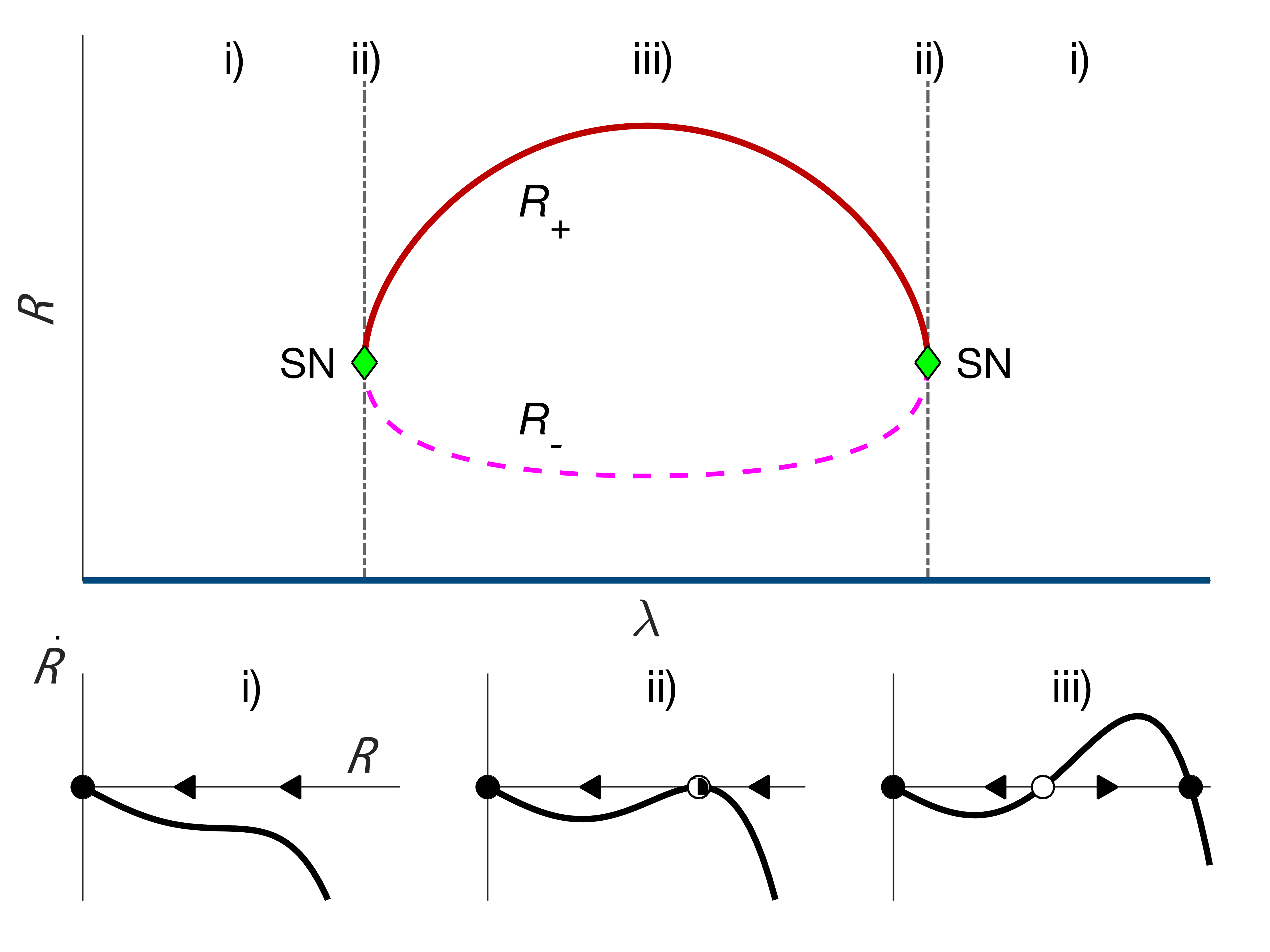}
\caption{ 
Upper panel: Steady-state solutions as a function of the bifurcation parameter $\lambda$. In our nanomechanical resonator $\lambda$ is a function of   $V_\mathrm{sd}$. Solid/dashed lines are stable ($R_+$)/unstable ($R_-$) solution branches. The isola  emerges as a result of non-monotonic nonlinear damping. 
Lower panel: sketches of the phase portrait of the vibration radius.  (i) The radial phase portrait for  $\lambda$ below and above the isola bifurcation point. The solid dot is a stable solution.  (ii) The saddle-node bifurcation (SN) at a finite vibration radius, indicating the onset of an isola.  
(iii) Bistable region with co-existing zero and large amplitude states. The open circle is an unstable  solution, which corresponds to an unstable limit cycle.
The arrows in panels (i), (ii), and (iii) indicate the dynamical flow.
 }
\label{fig:models} 
\end{figure} 

Since the vibration frequency is much higher than all other rates and frequencies in the system, the vibrations can be described in the rotating frame using a complex vibration amplitude $z(t) = C_z [q+ i(p/m\omega_0)]\exp(i\omega_0 t)$, where $q$
and $p$ are the mode coordinate and momentum, $m$ is its effective mass, $\omega_0$ is the eigenfrequency 
and 
$C_z$ is a scaling parameter. 
In the rotating wave approximation (RWA) the equation of motion (see Sec.~S3 of the   Supporting Information\cite{SI}) 
reads 
\begin{equation}
\dot{z}=-\left[ \Gamma +  (\gamma_\mathrm{nlf}-i\gamma_D)|z|^2+
 |z|^4 \right]z. 
\label{eq:mark}
\end{equation}
Here $\Gamma$ and $\gamma_\mathrm{nlf}$ are the  coefficients of linear and nonlinear friction, whereas $\gamma_D$ is the Duffing nonlinearity.  
The RWA applies  provided $|\dot z|\ll \omega_0|z|$. 
The right-hand side of Eq.~(\ref{eq:mark}) is \MD{our minimalistic model of nonlinear friction: it is} an expansion in $z$, valid when the vibration amplitude is comparatively small, so that the decay rate and the change of the vibration frequency are $\ll \omega_0$. 
\MD{The term  $\propto |z|^4z$} describes \MD{quintic}   nonlinear friction, cf.~\cite{strogatz2014nonlinear}. \MD{It}  must be taken into account where the conventional friction coefficients $\Gamma$ and $\gamma_\mathrm{nlf}$ become small. A simple  microscopic model of  such friction  is provided 
in   Supporting Information~S4 \cite{SI}. 
In distinction from the \MD{conventional} analysis \MD{of the onset of self-oscillations, which uses the model (\ref{eq:mark}) without the quintic term}, 
to describe the experiment we have to assume that the parameter $\Gamma$ remains positive, and it is $\gamma_\mathrm{nlf}$ that is the bifurcation parameter that changes sign.

We rewrite Eq.~\eqref{eq:mark} in polar coordinates by setting $z(t)=R{}(t) e^{i\theta(t)}$, where $R$ and $\theta$ are the vibration amplitude and the ``slow'' part of the vibration phase. From Eq.~\eqref{eq:mark} 
 \begin{equation}
\dot R=-f_\text{nlf} R,  \qquad  
\dot{\theta}=\gamma_D  R^2,   
\label{eq:radial}
\end{equation}
with    the coefficient of nonlinear friction being
\begin{equation}
f_\text{nlf}=\Gamma+\gamma_\mathrm{nlf} R^2+ R^4.
\label{eq:nff}
\end{equation}
\MD{For $\Gamma >0$ } the quiet state $R=0$ is stable.  
\MD{If}   $\Gamma$ becomes negative, the state $R=0$ becomes unstable, and \MD{for $\gamma_\mathrm{nlf}>0$} there emerges a  stable limit cycle with radius $\propto (|\Gamma|/\gamma_\mathrm{nlf})^{1/2}$ (supercritical Hopf bifurcation). If $\Gamma$ is positive but $\gamma_\mathrm{nlf}$ becomes negative,   there emerges an unstable limit cycle with radius $(\Gamma/|\gamma_\mathrm{nlf}|)^{1/2}$ (subcritical Hopf bifurcation).


The term $R^4$ in $f_\mathrm{nlf}$ leads to the onset of a stable limit cycle for $\gamma_\mathrm{nlf}<0$ and $\Gamma>0$ along with an  unstable one.  
The radii of the cycles as given by the condition $f_\mathrm{nlf}=0$ are
\begin{align}
    \label{eq:R+-}
    R_\pm = \frac{1}{\sqrt{2}} \left[-\gamma_\mathrm{nlf} \pm \sqrt{ \gamma_\mathrm{nlf}^2 - 4\Gamma}\right]^{1/2}
\end{align}
For $-\gamma_\mathrm{nlf}> 2\Gamma^{1/2}$ the cycle with the radius $R_+$ is stable, whereas the cycle with the radius $R_-$ is unstable. At $\gamma_\mathrm{nlf}=- 2\Gamma^{1/2}$ the two cycles merge and annihilate one another in a saddle-node bifurcation. For smaller $|\gamma_\mathrm{nlf}|/2\Gamma^{1/2}$ they disappear.

We now relate  model (\ref{eq:mark})  to the experiment. The values of  $\Gamma$ and $\gamma_\mathrm{nlf}$ change with the control parameter $V_\mathrm{sd}$. In particular, $\Gamma$ decreases as we approach the bistability region in Fig.~\ref{fig:observation}(b) \cite{Urgell2020}. 
The key observations are: (i) a stable zero-amplitude state and a stable limit cycle co-exist in a certain parameter range, (ii) the zero-amplitude state is stable  not only outside, but also inside this range, and (iii) the limit cycle is excited and collapses with the varying parameters while still having a large amplitude.  This scenario is qualitatively different from the standard subcritical Hopf bifurcation that is accompanied by hysteresis.

 A minimalistic picture that describes the experiment  is that, as the source-drain voltage $V_\mathrm{sd}$ varies, there first occurs a saddle-node bifurcation  at $-\gamma_\mathrm{nlf}= 2\Gamma^{1/2}$. At this bifurcation there emerge the stable and unstable limit cycle with   amplitudes $R_\pm$.   \MD{As $V_\mathrm{sd}$ varies further,} these limit cycles  merge together and disappear via another saddle-node bifurcation. This is illustrated in Fig.~\ref{fig:models}. 

To link the above phenomenology to Eqs.~\eqref{eq:radial} and \eqref{eq:nff} we should consider how the parameters of these equations depend on $V_\mathrm{sd}$. A major factor is the $V_\mathrm{sd}$-dependence of $\gamma_\mathrm{nlf}$, since this parameter becomes negative and, moreover, exceeds $2\Gamma^{1/2}$ in the absolute value. It should be noted that the linear friction coefficient $\Gamma$ also depends on $V_\mathrm{sd}$, as reported in \cite{Urgell2020}. Overall, \MD{$\gamma_\mathrm{nlf}$}   should be non-monotonic \MD{as a function of $V_\mathrm{sd}$} to allow for both the onset and the disappearance of the bistability with the increasing $V_\mathrm{sd}$.  
The analysis simplifies for the gate voltage $V_g$ where the range of $V_\mathrm{sd}$ in which the zero-amplitude state co-exists with the vibrational state is narrow. In this case one can approximate 
\begin{align}
    \label{eq:small_range}    \
    \gamma_\mathrm{nlf}(V_\mathrm{sd}) + 2\sqrt{\Gamma(V_\mathrm{sd})} = -\eta (V_B^{(1)}-V_\mathrm{sd})(V_\mathrm{sd}-V_B^{(2)})
\end{align}
Here, the parameter $\eta>0$ is a scaling parameter and  $V_B^{(1,2)}$ are the  bifurcational values of $V_\mathrm{sd}$. 
The CNT shows bistabilty in the range  $V_B^{(2)}<V_\mathrm{sd}< V_B^{(1)}$. 

\begin{figure}[htb!]
\centering 
\includegraphics[width=0.65\textwidth]{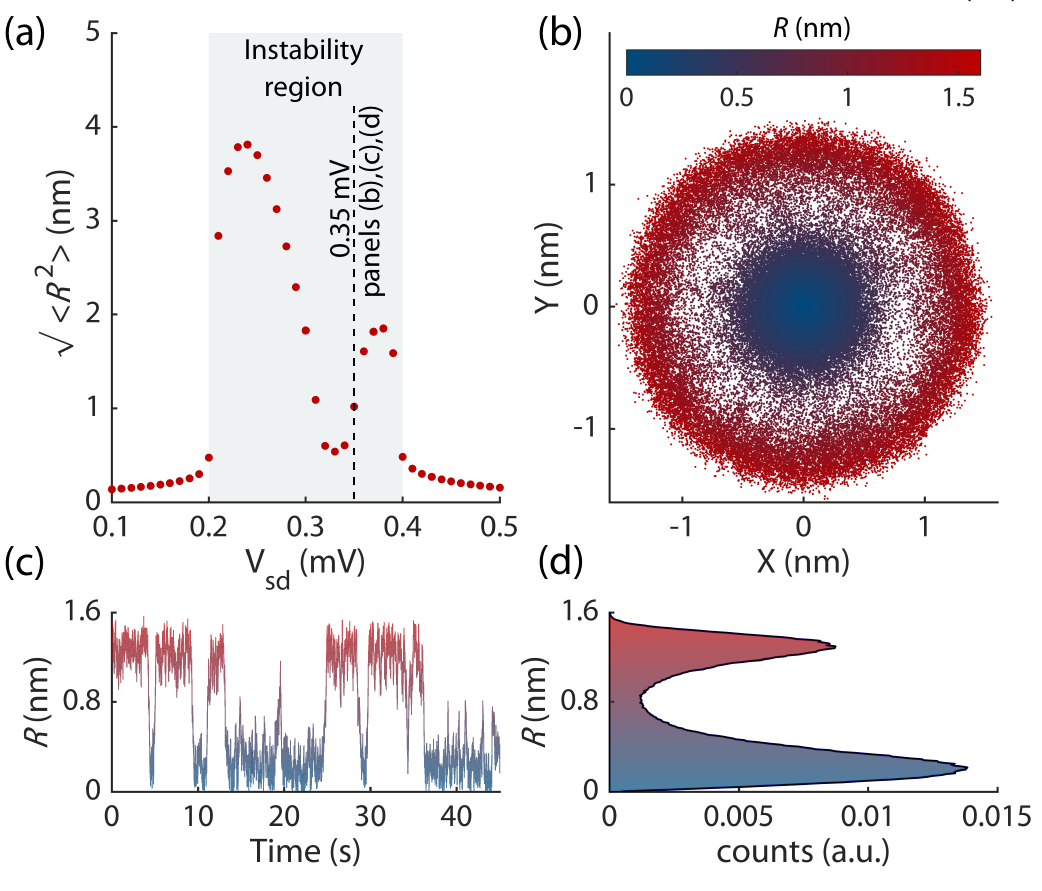} 
\caption{   Numerical simulations of the fluctuation dynamics  of the nanotube. 
(a)  
The  standard deviation  of the nanotube displacement $\sqrt{<R^2>}$ as a function of the source-drain voltage $V_\text{sd}$. Each point is the average of ten simulations. 
The stochastic dynamics is characterized for one simulation at $V_\text{sd}= $\SI{0.35}{\milli\volt} (dashed line) in panels (b)-(d). 
(b) The phase space of the two quadratures of the motion (X,Y). 
 (c) Fluctuations of the amplitude $R=\sqrt{\text{X}^2+\text{Y}^2}$ in time. 
 (d) The amplitude  histogram  of the time trace of panel (c)  normalized with respect to the total number
of observations. 
 Details of the simulations of the stochastic dynamics of the nanotube are provided in Sec.~S5 of the   Supporting Information \cite{SI}.
 }
\label{fig:traj} 
\end{figure}

The quadratic dependence of  $\gamma_\mathrm{nlf}$  and the corresponding dependence of $R_+$  on $V_\mathrm{sd}$   provide an insight into  our observations, but do not fully describe the evolution of the dynamics within the instability region.
The complicated dependence of $R_+$ on $V_\mathrm{sd}$    can have several causes, \MD{including defects in the CNT that lead to a nonuniform electron density,  as well as the interplay of the Kondo effect and the Coulomb blockade, which depend on the bias and the gate voltage, thus modifying the tunneling and the polaronic effect and ultimately the friction force.} 
In Fig.~\ref{fig:traj} we show that the complex behavior of the vibration amplitude within the instability region can be effectively captured by considering a   {\corr non-monotonic} dependence of the nonlinear friction force on the source-drain voltage. Given that the dependence of the root mean square vibration amplitude $\braket{R^2}^{1/2}$ on $V_\mathrm{sd}$ in Fig.~\ref{fig:observation}(b) resembles an inverted quartic parabola, we describe the dynamics within the bistability region by a constant linear friction and a 5-parameter nonlinear friction, $\gamma_\mathrm{nlf}= \sum_{n=0}^4 (\gamma_n V_\mathrm{sd}^n)$.   To numerically obtain the parameters that match the experiment we positioned the saddle-node points of the isola at the boundaries of the instability region observed experimentally and constrained the amplitude of the self-oscillations to match the measured $\sqrt{<R^2>}$ values. 


We plot in Fig.~\ref{fig:traj}(a) the evolution of the amplitude variance as a function of $V_\text{sd}$ by simulating the stochastic dynamics of Eq.~\eqref{eq:mark} in the quadrature space,
in which each quadrature is affected by an independent random Wiener process (See Sec.~S5 of the   Supporting Information \cite{SI} for more details). \MD{The noise, which is  white in the slow time $\sim \Gamma^{-1}, (|\gamma_\mathrm{nlf}| \braket{R^2})^{-1}$, comes from different intrinsic sources, such as  hopping of the electrons on and off the CNT and creation and annihilation of thermal phonons nonlinearly coupled to the mode.}
In the  phase-space of the  two quadratures (X,Y)   \MD{the trajectories fluctuate about a circle with radius $\braket{R^2}^{1/2}$ or about the quiet state $R=0$, switching between them, see Fig.~\ref{fig:traj}(b) and (c). Respectively,  the stationary probability distribution of $R(t)$ displays two peaks, as seen in Fig.~\ref{fig:traj}(d).} 

We emphasize that the {\corr non-monotonic} dependence of \MD{the nonlinear friction parameter $\gamma_\mathrm{nlf}$}   on $V_\mathrm{sd}$\MD{, with $\gamma_\mathrm{nlf}$ being negative in a certain range of $V_\mathrm{sd}$,}    is critical for the emergence of the isola   and the hysteresis-free bistability\MD{. At the same time, fluctuations in the system are}  crucial \MD{for revealing}   the bistability.

Although  isolas in  multistable systems  have attracted much attention   theoretically 
\cite{Habib2018,Hill2016,Cirillo2017,Melot2024}, their experimental demonstrations have almost exclusively been limited to macroscale systems under periodic driving  \cite{Bureau2014,Gatti2017,Detroux2018}. 
In mesoscopic systems, the only reported observation of isolas involves forced vibrations of a  nonlinear microresonator with coupled vibrational modes\cite{Dong2018}.  \MD{In nanomechanis, isolas have not been observed.
}  
Here, we demonstrate the existence of an isolated vibrational state in a \MD{nanomechanical} system subject to a time-independent drive. 
Specifically, we show that driving a carbon nanotube by a dc source-drain voltage $V_\mathrm{sd}$ leads to the onset of bistability, in which a stable quiet state coexists with periodic self-sustained vibrations. \MD{We show that the bistability is non-hysteretic: varying the control parameter would not lead to switching between the branches of the stable states.} 
{\corr We} \MD{also} provide a \MD{minimalistic phenomenological model that describes the effect and indicate the mechanisms that can underlie this model.} 
 




Supporting Information:  Additional analysis of the statistics of stochastic switching of the carbon nanotube,  discussion on the friction force generated due to electrothermal backaction, theoretical description of the mechanism leading to quintic nonlinear friction, and more information on the numerical simulations.

\begin{acknowledgement} 
\MD{We are grateful to Fabio Pistolesi for the discussion of the electron-phonon coupling in CNT.} Financial support was provided from the European Union’s research and innovation programme under ERC starting grant no. 802093, ERC consolidator grant no. 101125458, and ERC advanced grant  no. 692876. 
MID acknowledges partial support  from the US Defense Advanced Research Projects Agency (Grant No. HR0011-23-2-004) and from the Moore Foundation (Grant No. 12214).  
PB acknowledges partial support from the European Union’s NextGenerationEU programme, in the framework of PRIN 2022, project DIMIN. 
AB acknowledges MICINN Grant No. RTI2018-097953-B-I00 and PID2021-122813OB-I00, AGAUR (Grant No. 2017SGR1664), the Fondo Europeo de Desarrollo, the Spanish Ministry of Economy and Competitiveness through Quantum CCAA, TED2021-129654B-I00, EUR2022-134050, and CEX2019-000910-S [MCIN/AEI/10.13039/501100011033], MCIN with funding from European Union NextGenerationEU (PRTR-C17.I1) and Generalitat de Catalunya, CERCA, Fundacio Cellex, Fundacio Mir-Puig. 
\end{acknowledgement} 


 \bibliography{biblio}
\end{document}